\documentstyle[twocolumn,prb,aps,epsfig,subeqnar]{revtex}

\begin{document}

\title{Stability of stationary states in the cubic nonlinear Schr\"odinger equation: applications to the Bose-Einstein condensate}

\author{Lincoln D. Carr$^{1}$\cite{byline}, J. Nathan Kutz$^{2}$, William P. Reinhardt$^{1,3}$\\}
\address{$^{1}$Department of Physics, University of Washington, Seattle, WA 98195-1560, USA\\}
\address{$^{2}$Department of Applied Mathematics, University of Washington, Seattle, WA 98195-2420, USA\\}
\address{$^{3}$Department of Chemistry, University of Washington, Seattle, WA 98195-1700, USA\\}

\date{\today}

\maketitle
\begin{abstract}
The stability properties and perturbation-induced dynamics of the full set of stationary states of the nonlinear Schr\"odinger equation are investigated numerically in two physical contexts:  periodic solutions on a ring and confinement by a harmonic potential.  Our comprehensive studies emphasize physical interpretations useful to experimentalists.  Perturbation by stochastic white noise, phase engineering, and higher order nonlinearity are considered.  We treat both attractive and repulsive nonlinearity and illustrate the soliton-train nature of the stationary states.
\end{abstract}

\section{Introduction}
\label{sec:intro}

The one-dimensional nonlinear Schr\"odinger equation (NLSE) is ubiquitious.  Among other natural phenomena, it models the dilute-gas Bose-Einstein condensate (BEC) in the quasi-one-dimensional regime~\cite{carr22}, light pulses in optical fibers~\cite{hasegawa1}, Bose-condensed photons~\cite{ciao1}, helical excitations of a vortex line~\cite{hasimoto1}, and spin waves in magnetic materials~\cite{kalinikos1}.  In this article application of the NLSE to the BEC is emphasized.  As current~\cite{ketterle1,matthews1,williams1} and proposed~\cite{close1} BECs are studied in traps of differing topologies, issues of stability for periodic solutions on a ring and for confinement in a harmonic potential are both of interest.

In a recent pair of articles~\cite{carr15,carr16} the full set of periodic solutions of the stationary NLSE on a ring was presented in closed analytic form.  Here we extend our previous work by numerically studying the stability of such stationary states in response to initial and continuous stochastic white noise, phase engineering, and higher order nonlinearity, and by undertaking a similiar set of studies for the harmonic potential.  In the BEC stochastic white noise arises from the effect of the thermal cloud of uncondensed atoms on the condensate, and is nominally a controllable parameter; simple phase profiles are already in use in experiments~\cite{denschlag1,burger1}; and higher order nonlinearity models both mean field effects beyond the Gross-Pitaevskii approximation~\cite{pitaevskii1,gross1} and reduced dimensionality~\cite{kolomeisky1}.

As has been shown previously~\cite{carr15,carr16}, periodic stationary states on a ring can be characterized as bright or dark soliton trains for attractive or repulsive nonlinearity, respectively.  It follows that our numerical studies can in part be understood by combining the extensive analytical results from the physical and mathematical literature on the stability of single solitons~\cite{elgin1,kivshar1} and the nature of two-soliton interactions~\cite{zakharov1,zakharov2,lerner1,gordon1} and extrapolating to soliton-trains.  However, because we work with bounded systems it is necessary to consider scaling issues.  Adding an external potential term to the NLSE adds an additional parameter which cannot be scaled into space, time, or normalization of the amplitude.  This extra parameter can be incorporated as the coefficient of the nonlinearity.  In the context of the BEC, it has a simple physical interpretation as being proportional to the number of atoms in the condensate.  Equivalently it can be incorporated as the coefficient of the dispersion, in which case it has a physical interpretation as $\xi^2$, where $\xi$ is the healing length which measures the length scale of changes in the condensate wavefunction~\cite{dalfovo1}.  As a result there are three different scale regimes: \emph{overlapping}, \emph{adjacent}, and \emph{well-separated}.  These terms, which apply to the spacing of solitons, will be defined in Sec.~\ref{ssec:review}.  The stability properties and perturbation-induced dynamics of soliton-trains in these three regimes can differ greatly.

The manner in which periodic soliton-train stationary-states on a ring respond to stochastic white noise and higher order nonlinearity carries directly over to the harmonic potential.  We present a representative set of stationary states for the harmonic potential and show how they can be manipulated via simple phase profiles, or \emph{phase engineered}.  It is also shown how the form of the ground state depends on the strength of the nonlinearity, i.e. the number of atoms.  As single dark or grey solitons have recently been observed for the first time in the BEC in a harmonic potential~\cite{denschlag1,burger1}, our investigations are of immediate benefit to BEC experiments.  

We especially focus on the physical meaning of different types of stability.  As the NLSE has no dissipation, solutions to it are at best stable, but not asymptotically so~\cite{hirsch1}.  Furthermore, stability on the time scale of an experiment is not the same as mathematical stability.  For repulsive nonlinearity stationary states are generally stable; for attractive nonlinearity, they are \emph{quasi-periodically stable}, as shall be defined in Sec.~\ref{sssec:att}.  In the right parameter regimes such states can be both observable and useful for experimentalists.

\section{Periodic solutions on the ring}
\label{sec:pbc}

The analytic description of all periodic stationary states of the 1D NLSE on a ring of circumference $L$ is reviewed.  The dynamics in response to perturbation by stochastic white noise and higher order nonlinearity are then illustrated and discussed.

\subsection{Review of analytic solutions}
\label{ssec:review}

The NLSE in 1D has a number of special properties which are described in the mathematical literature.  It is integrable~\cite{sulem1,kivshar3}, may be solved exactly by the inverse scattering transform~\cite{zakharov1,zakharov2} and has a countably infinite number of conserved quantities~\cite{miura1,miura2}.  We have taken advantage of its special properties in finding its stationary solutions.  It models the BEC in the quasi-one-dimensional regime, which is obtained when the transverse dimensions of the condensate are on the order of its healing length and its longitudinal dimension is much longer than its transverse ones~\cite{carr15}.

We begin with the cubic NLSE with an additional scaling factor which incorporates the strength of the nonlinearity:
\begin{subeqnarray}
& [-\partial_{x}^{2}\:\pm\eta\mid\!f(x,t)\!\mid^{2}\,]\,f(x,t) = \imath\partial_{t}\,f(x,t)\, , \\
& [-\xi^2\partial_{x}^{2}\:\pm\mid\!f(x,t)\!\mid^{2}\,]\,f(x,t) = \imath\partial_{t}\,f(x,t)\, .
\label{eqn:nlsetime}
\end{subeqnarray}
The two forms of Eq.~\ref{eqn:nlsetime} have different physical emphases.  In the first, $\eta$ is proportional to the number of condensed atoms $N$ and the s-wave scattering length between atoms; in the second, $\xi$ is the healing length~\cite{dalfovo1}, which gives the length scale of variations in the condensate wavefunction $f(x,t)$.  As $\eta\propto\xi^{-2}$, the two forms are equivalent.  The $\pm$ refers to repulsive or attractive two-body atomic interactions, respectively, and $x$ and $t$ are rescaled space and time.

\begin{figure}
\begin{center}
\epsfig{figure=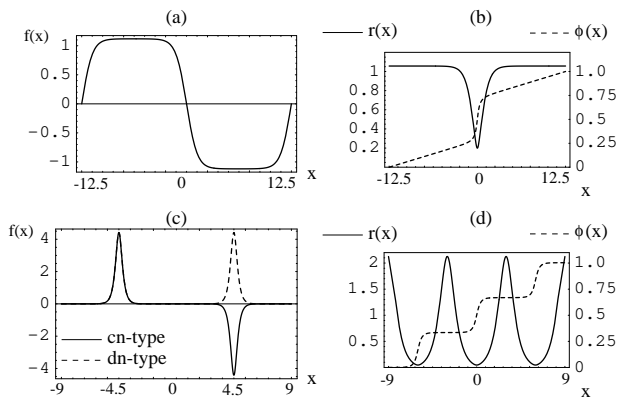,width=8.2cm}
\end{center}
\caption{Shown are the five periodic stationary solution-types in one dimension.  (a) Amplitude of real solution and (b) amplitude and phase of intrinsically complex solution for repulsive nonlinearity; (c) amplitude of dn-type and cn-type solutions and (d) amplitude and phase of intrinsically complex solution for attractive nonlinearity.}
\label{fig:5solns}
\end{figure}

Assuming the wavefunction to be of the form $f(x,t)=f(x)e^{-\imath\mu t}$ results in the stationary NLSE
\begin{equation}
[-\partial_{x}^{2}\:\pm\eta\mid\!f(x)\!\mid^{2}\,]\,f(x) = \mu\,f(x)\, ,
\label{eqn:nlse}
\end{equation}
where $\mu$ is the eigenvalue which corresponds to the chemical potential of the BEC.  All stationary solutions to Eq.~(\ref{eqn:nlse}) may be written in terms of Jacobian elliptic functions~\cite{bowman1}.  The properties of such functions are reviewed elsewhere~\cite{carr15,bowman1,abramowitz1}.  There are five normalizable, symmetry-breaking, periodic solution-types to the stationary NLSE on the ring.  They are pictured in Fig.~\ref{fig:5solns}.  All five solution-types are found by solving Eq.~(\ref{eqn:nlse}) subject to normalization and boundary conditions, and are described in detail in references [10] and [11].

Real stationary states in one-to-one correspondence with those of the particle-on-a-ring problem in linear quantum mechanics are

\begin{equation}
f(x)=A\,{\rm sn}\left(\frac{2j K(m) x}{L}+\delta\,\Big|\,m\right)
\label{eqn:sn}
\end{equation}
and
\begin{equation}
f(x)=A\, {\rm cn}\left(\frac{2j K(m) x}{L}+\delta\,\Big|\, m\right)
\label{eqn:cn}
\end{equation}
for repulsive and attractive nonlinearity, respectively.  They shall be designated \emph{sn-type} and \emph{cn-type}, in accord with their functional form.  For the case where $j=2$, Eq.~(\ref{eqn:sn}) is shown in Fig.~\ref{fig:5solns}(a) and Eq.~(\ref{eqn:cn}) in Fig.~\ref{fig:5solns}(c).  $A$ is the amplitude; $j\in\{2,4,6,\cdots\}$ is the number of nodes; $K(m)$ is the complete Jacobian elliptic integral which is the quarter period of these functions; $\delta$ is an arbitrary translational offset, which leads to a Kosterlitz-Thouless-type entropy in one dimension~\cite{kosterlitz1}; $L$ is the circumference of the ring; and $0\leq m\leq 1$ is the Jacobian elliptic parameter.  The general notation sn$(u\mid m)$ is standard for Jacobian elliptic functions~\cite{bowman1,abramowitz1}.

The Jacobian elliptic parameter $m$ governs the strength of the nonlinearity $\eta$.  As $m\rightarrow 0^+$, $\eta\ll 1$, sn $\rightarrow$ sin, and cn $\rightarrow$ cos, respectively.  This is the linear, sinusoidal limit which reproduces the particle-on-a-ring stationary solutions from linear quantum mechanics.  As $m\rightarrow 1^-$, $\eta\gg 1$, sn $\rightarrow$ tanh, and cn $\rightarrow$ sech.  These are the dark and bright soliton, stationary solutions to the NLSE on the infinite line, respectively~\cite{zakharov1,zakharov2}.  This shows the connection between these stationary solutions and solitons.  On the ring the $j^{{\rm th}}$ stationary state is a $2j$ soliton-train.

In addition to the above two solution-types there are three nodeless, symmetry-breaking solution-types which have no analogue in the particle-on-a-ring problem in linear quantum mechanics.  The first, a solution for attractive nonlinearity, is real:

\begin{equation}
f(x)=A\, {\rm dn}\left(\frac{2j K(m) x}{L}+\delta\,\Big|\, m\right)\, ,
\label{eqn:dn}
\end{equation}
where $j\in\{1,2,3,\cdots\}$.  It shall be designated \emph{dn-type}.  An example is shown in Fig.~\ref{fig:5solns}(c).  The other two types are intrinsically complex.  $f(x)\equiv r(x)\exp(\imath\phi(x))$ and for repulsive nonlinearity

\begin{equation}
r(x)^{2}=A^2\left(1-\gamma\,{\rm dn}^{2}\left(\frac{2jK(m) x}{L} + \delta \,\Big|\, m \right) \right)
\label{eqn:repcom}
\end{equation}
while for attractive nonlinearity
\begin{equation}
r(x)^{2}=A^2\left({\rm dn}^{2}\left(\frac{2jK(m) x}{L} + \delta\,\Big|\, m\right)-\gamma(1-m)\right)\, ,
\label{eqn:attcom}
\end{equation}
where in both cases the phase must be found by numerical integration from the equation

\begin{equation}
\phi^{\prime}(x)=\frac{\alpha}{r(x)^2}\, .
\label{eqn:comphase}
\end{equation}
The phase and amplitude for $j=1$ and $j=3$ are shown in Fig.~\ref{fig:5solns}(b) and~\ref{fig:5solns}(d).  In the repulsive case $A^2\gamma$ is the depth of the density minima below the constant background.  When $\gamma = 1$ Eq.~(\ref{eqn:sn}) is recovered.  In the attractive case $\gamma$ interpolates between the cn-type and dn-type solutions in Eqs.~(\ref{eqn:cn}) and~(\ref{eqn:dn}).  In both cases $0\leq\gamma\leq1$; $\alpha$ is a constant of integration; $j\in\{1,2,3,\cdots\}$ is the number of density minima or maxima, respectively; and each stationary state has a complex-conjugate, degenerate partner.

The intrinsically complex solutions for repulsive nonlinearity are interpreted as density-notch solitons moving with speed $c$ on the ring with an opposing momentum boost of the condensate of speed $-c$, which results in a stationary state in the lab frame.  Density-notch solitons have a speed between zero and the Boguliubov sound speed, ranging from maximal to zero depth, respectively~\cite{reinhardt1}.  Those not of maximal depth are called grey solitons, while those which are of maximal depth and therefore form a node are called dark solitons.  Fig.~\ref{fig:5solns}(b) shows the bounded, quantized version of a stationary grey soliton on a ring.

All attractive symmetry-breaking, longitudinally periodic, stationary solution-types, i.e. the cn and dn-types shown in Fig.~\ref{fig:5solns}(b) and the intrinsically complex one shown in Fig.~\ref{fig:5solns}(d), are described by the $C_j$ point symmetry group, where $j$ is the number of peaks.  There are $j$ nearly degenerate solutions.  For even $j$ there is a real dn-type-cn-type pair and $(j-2)/2$ degenerate intrinsically complex pairs.  For odd $j$ there is a real dn-type solution and $(j-1)/2$ degenerate intrinsically complex pairs.  Thus by group theory these three solution-types form the complete set of periodic stationary states made of evenly spaced peaks.

The existence of nodeless solutions requires a minimum $\eta$~\cite{carr15,carr16}.  Based on this bound, the effective spatial extent of a soliton is $2\pi\sqrt{6}\,\xi$ or $2\pi\sqrt{2}\,\xi$ for repulsive or attractive nonlinearity.  In our plots we shall choose $\xi=1$ and let the length of the ring determine the strength of the nonlinearity, since on a finite interval the important quantity is $\xi/L$.  The three scale regimes mentioned in the introduction are then defined as follows:  $j\xi/L\ll 1$ is the \emph{well-separated} regime; $j\xi/L\sim 1$ is the \emph{adjacent} regime; and $j\xi/L\gg 1$ is the \emph{overlapping} regime.

\subsection{Dynamics under perturbation}
\label{ssec:periodic}

Of experimental interest is the stability of the above-mentioned stationary states on physical time scales.  This is investigated numerically in this and the following sections.  The algorithm employed uses a fourth order Runge-Kutta in time and a filtered pseudospectral method in space to propagate solutions over space-time intervals of interest.  Although some solutions are illustrated over short time scales for presentation purposes, all simulations were performed over time scales longer than experimental lifetimes of the BEC~\cite{carr22}.  In this section white noise and higher order nonlinearity are considered; phase engineering of periodic solutions on the ring has been investigated elsewhere~\cite{carr22}.  Throughout our presentation the level of initial white noise will represent a number of uncondensed atoms on the order of $1\%$ of the condensed ones~\cite{whiteNoise}.

Kolomeisky {\it et al.}~\cite{kolomeisky1} showed that for the BEC a true one-dimensional mean field theory, rather than the quasi-one-dimensional regime of the three-dimensional NLSE, has quintic rather than cubic nonlinearity.  For this reason a quintic perturbation is chosen for higher order nonlinearity.  Transverse confinement on the order of the healing length should be described by the NLSE, while transverse confinement on the order of the atomic interaction length should be described by a one-dimensional mean field theory.  As these scales are two or three orders of magnitude apart, the combination of cubic and quintic nonlinearity may describe the intermediate regime.  This full range of confinement scales should soon be available for experimental investigation~\cite{key1}.

\subsubsection{Repulsive nonlinearity}
\label{sssec:rep}

In response to white noise, density-notch soliton-trains drift but are otherwise stable.  This may be understood by extrapolation from the stability properties of single solitons and two-soliton interactions.  Density notch solitons repel elastically.  Single solitons respond to stochastic perturbation by emission of radiation and by a change in velocity and depth.  Velocity and depth are a function of a single parameter, sometimes called the soliton phase angle.  A full analysis requires separate consideration of the constant background and the density notch.  Kivshar and Yang~\cite{kivshar1} have completed this analysis for single solitons on the infinite line by using a variational approach.

\begin{figure}
\begin{center}
\epsfig{figure=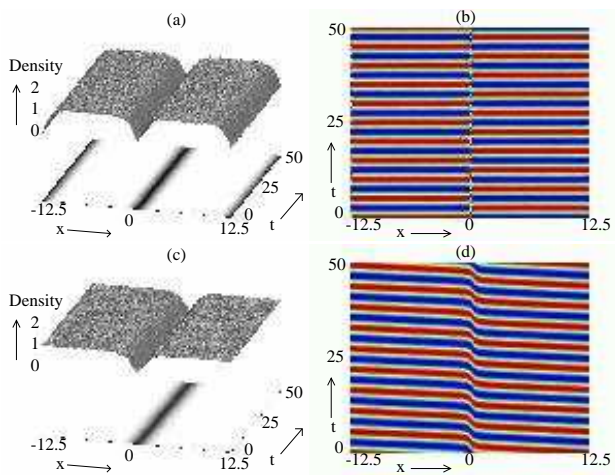,width=8.2cm}
\end{center}
\caption{
Stability for repulsive nonlinearity, periodic solutions on the ring.  (a) Density and (b) phase of real, sn-type solution with initial white noise; (c) density and (d) phase of intrinsically complex solution.  The lowest energy solution is shown for each type.  A short time scale was used to illuminate the phase but the same stability properties hold over an order of magnitude longer time.  The phase is plotted mod $2\pi$.
}
\label{fig:2}
\end{figure}

In Fig.~\ref{fig:2}(a)-(b) the evolution of a sn-type stationary state described by Eq.~(\ref{eqn:sn}) with initial white noise is shown.  This is the lowest such state on the ring.  On a longer time scale there is some drift.  In this case the two density notches retain their spacing and phase relationship, which shows that, as with single solitons, only the velocity of the soliton train changes.  In the box this same stationary state is the first excitation above the ground state.  Numerical studies, though not shown here, reveal the single density notch in the center drifts randomly between the box boundaries until it comes close to one, at which point it reverses direction abruptly.  This can be explained by interpreting the falling off of the wavefunction at the boundaries as pinned solitons~\cite{reinhardt1}, which repel the central density notch when it comes within the interaction length of $\pi\sqrt{6}\,\xi$.

In Fig.~\ref{fig:2}(c)-(d) the evolution of an intrinsically complex stationary state described by Eqs.~(\ref{eqn:repcom}) and~(\ref{eqn:comphase}) with initial white noise is shown.  This is the lowest symmetry-breaking state on the ring.  It consists of both a background phase ramp and a density notch, and is quite stable.  Because phase quantum number is conserved for the NLSE, the background cannot be perturbed into a lower energy state.  And a grey density notch has the same response to perturbation as a dark density notch.  Therefore the two together emit radiation and, on longer time scales, drift, but remain otherwise unchanged. 

\begin{figure}
\begin{center}
\epsfig{figure=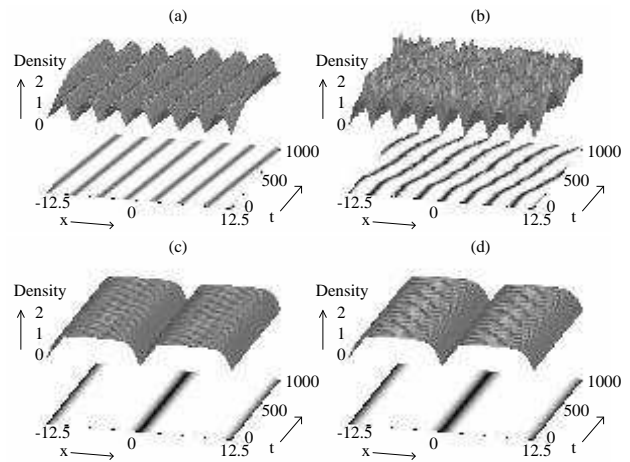,width=8.2cm}
\end{center}
\caption{
More stability properties for repulsive nonlinearity.  (a) Highly excited intrinsically complex solutions are stable to perturbation by initial white noise, despite slow drift.  (b) Continuous noise builds up because there is no dissipation term in the NLSE, but even highly excited solutions remain stable.  For (c) positive and (d) negative quintic nonlinearity without white noise the stationary states develop slight fluctuations but remain stable on very long time scales.
}
\label{fig:3}
\end{figure}

In Fig.~\ref{fig:3}(a) the evolution of a highly excited state is shown.  This demonstrates that for repulsive nonlinearity there is no difference between the stability properties of soliton-train stationary states in the overlapping, adjacent, or well-separated regimes, a fact not \emph{a priori} apparent from single soliton considerations.  In Fig.~\ref{fig:3}(b) continuous white noise is added.  This perturbs the equation rather than the initial solution.  As there is no dissipative term in the NLSE, the noise builds up over 1000 time units to a very high level.  Even in this case, the only instability exhibited by a highly excited, intrinsically complex solution is drift.

Finally, in Fig.~\ref{fig:3}(c)-(d) quintic nonlinearity is added to the same sn-type stationary state as was used in Fig.~\ref{fig:2}(a)-(b).  Eq.~(\ref{eqn:nlsetime}) then becomes:

\begin{equation}
[-\partial_{x}^{2}\:\pm\eta\mid\!f(x,t)\!\mid^{2}+b\mid\!f(x,t)\!\mid^{4}\,]\,f(x) = \imath\partial_t f(x,t)
\label{eqn:nlsequintic}
\end{equation}
We used $b=\pm0.1$ for (c) and (d), respectively.  In both cases the initial state was found to be stable with respect to this perturbation, with only small modulations in the background and a slight reshaping of the density notch~\cite{kapitula1}.  For intrinsically complex stationary states a similiar set of results was obtained, except that the balance between the constant background and the grey density notch was no longer the same, so that the solution translated.  This suggests that soliton-train stationary states may exist for transverse confinement of the BEC on scales intermediate between the atomic interaction length and the healing length.

\subsubsection{Attractive nonlinearity}
\label{sssec:att}

The study of soliton trains differs from that of single solitons or two-soliton interactions because it no longer suffices to consider the well-separated limit.  The exact equation for interactions between two bright solitons has been derived in full by Gordon~\cite{gordon1}.  As it is difficult to interpret when the solitons are not resolvable, he has considered the non-overlapping limit and found 

\begin{eqnarray}
\ddot{q}=-4\exp(-2q)\cos(2\Psi)\nonumber\\
\ddot{\Psi}=4\exp(-2q)\sin(2\Psi)
\label{eqn:twosoliton}
\end{eqnarray}
where $2q$ is the separation of the two solitons and $2\Psi$ is their relative phase.  The interaction depends exponentially on the separation and sinusoidally on the relative phase.  Desem and Chu~\cite{desem1} used this work to evaluate interaction minimization schemes in the context of fiber optics.

\begin{figure}
\begin{center}
\epsfig{figure=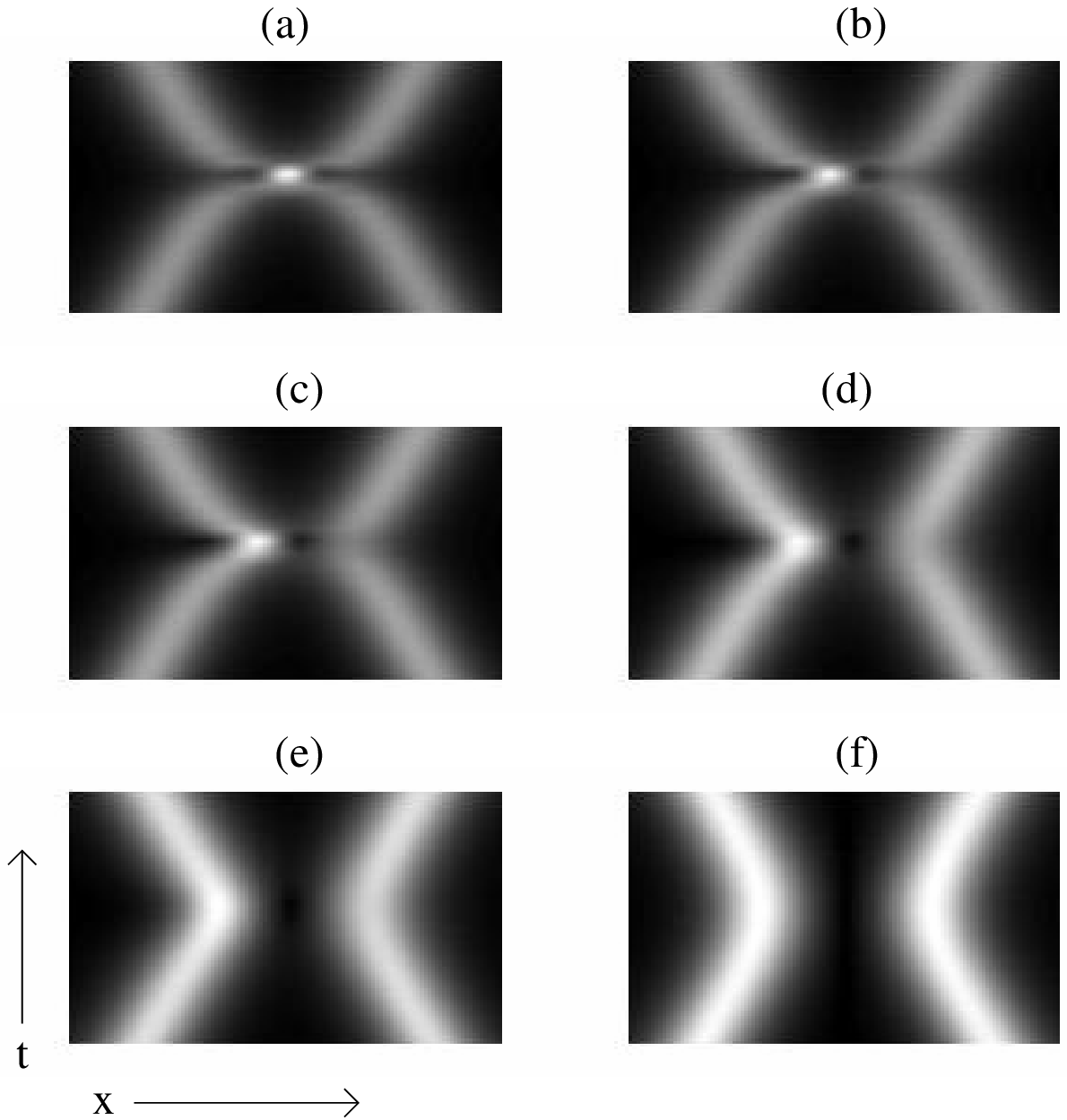,width=8.2cm}
\end{center}
\caption{
Phase dependence of bright soliton interactions.  Shown are the space-time projection of the density of colliding bright solitons.  The phase difference between the two solitons is (a) 0, (b) $\pi/16$, (c) $\pi/8$, (d) $\pi/4$, (e) $\pi/2$, and (f) $\pi$.  The time scale in each case is 50 time units and the interaction length is $2\pi\sqrt{2}$.  Note that the circumference of the ring is much greater than the length scale shown here.
}
\label{fig:4}
\end{figure}

We found that the key to understanding the stability of intrinsically complex stationary states perturbed by initial white noise for attractive nonlinearity lay in the overlapping regime of these two soliton interactions.  Figure~\ref{fig:4} shows the results of our studies.  Following the six subpanels of the figure from left to right and top to bottom, it is apparent that the topology of the spacetime profile changes from connected to disconnected as the phase varies from $0$ to $\pi$.  A hole opens in the center, rotates, and then separates the density profiles of the two solitons.  Thus the interaction changes continuously from completely attractive to completely repulsive.  For $2\Psi\in[\pi,2\pi]$ the topology reverses itself in the same way.

Single bright solitons respond to stochastic perturbation not only by emission of radiation and by a change in velocity, but also by a drift in phase~\cite{elgin1}.  As a soliton train requires a fixed phase relationship between its components, it is to be expected that the stability properties of solitons trains for attractive nonlinearity differ from those of repulsive nonlinearity, as may be seen for example in the numerical studies of densely-packed solitons by Arbel~\cite{arbel1}.

\begin{figure}
\begin{center}
\epsfig{figure=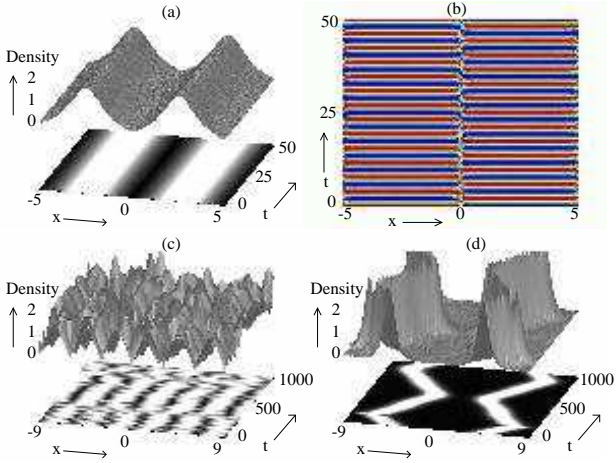,width=8.2cm}
\end{center}
\caption{
Stability and the importance of scale, attractive nonlinearity. (a) Density and (b) phase of the cn-type solution with initial white noise in the adjacent regime.  The phase is plotted mod $2\pi$.  However, in the (c) overlapping and (d) well-separated regimes the perturbed cn-type solution is unstable.
}
\label{fig:5}
\end{figure}

In Fig.~\ref{fig:5}(a)-(b) the evolution of cn-type stationary state described by Eq.~(\ref{eqn:cn}) is shown with initial white noise.  In the adjacent regime this solution is quite stable out to very long time scales.  However, in the overlapping regime such a stationary state undergoes interactions, while in the well-separated regime the peaks behave as individual solitons, as may be seen in Fig.~\ref{fig:5}(c)-(d).  The drift in phase is especially evident in (d), where in the first interaction there is clearly density exhange between the two solitons, despite their initial phase difference being $\pi$.

\begin{figure}
\begin{center}
\epsfig{figure=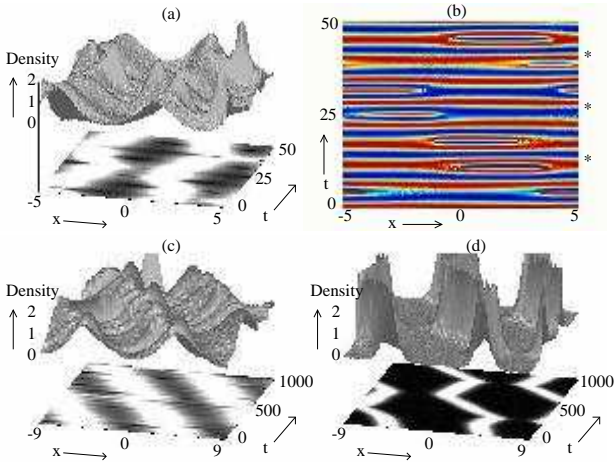,width=8.2cm}
\end{center}
\caption{
Quasi-periodic stability and the importance of scale, attractive nonlinearity. (a) Density and (b) phase of the nodeless, dn-type solution with initial white noise in the adjacent regime.  The asterisks in (b) mark the recurrence of the solution, clearly visible here on a short time scale.  (c) On longer time scales but still in the adjacent regime, the peaks exchange mass and drift, but continue to recur.  (d) In the well-separated regime the density peaks behave as independent solitons.  Although not shown here, the perturbed dn-type solution is also unstable in the overlapping regime.
}
\label{fig:6}
\end{figure}

In Fig.~\ref{fig:6}(a)-(b) the evolution of a real dn-type stationary state described by Eq.~(\ref{eqn:dn}) is shown with initial white noise in the adjacent regime.  As the phase difference between peaks is zero, it is not surprising that the solution seems to quickly go unstable.  However, there are quasi-periodic recurrences, which are noted with asterisks in (b).  For this reason the solutions are termed quasi-periodically stable.  This is especially evident on the longer time scale shown in Fig.~\ref{fig:6}(c), where it may be seen that though the peaks continue to exchange mass the overall integrity of the soliton-train remains intact.  The drift of the train is similiar to that found for the cn-type stationary states in the adjacent regime, and as before, in the overlapping regime it is unstable while in the well-separated regime the peaks behave as independent solitons.

\begin{figure}
\begin{center}
\epsfig{figure=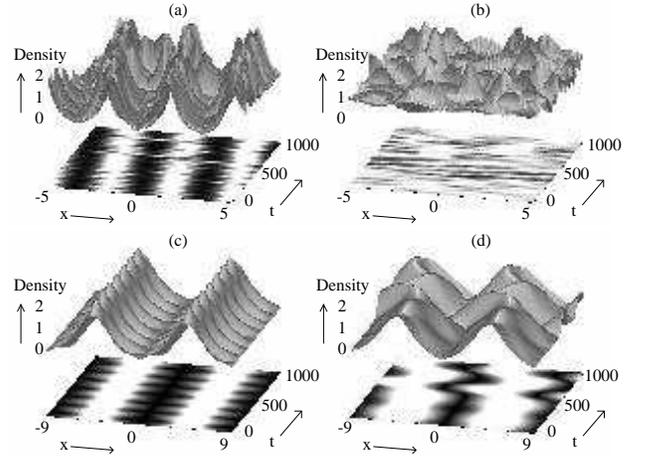,width=8.2cm}
\end{center}
\caption{
More stability properties for attractive nonlinearity.  (a) The long time behaviour of intrinsically complex solutions in the adjacent regime when perturbed by initial white noise is the same as for dn-type solutions.  (b) In the overlapping regime, intrinsically complex solutions are unstable but do not collapse to a ground state, due to conservation of phase quantum number.  Positive quintic nonlinearity causes the peaks to pulse while (d) negative quintic nonlinearity leads to large oscillations but not actual breakup of the solitons.}
\label{fig:7}
\end{figure}

The intrinsically complex stationary states have similiar properties to the real dn-type stationary states: in the adjacent regime they evolve and retain their overall integrity, and in other regimes they are unstable, as is shown in Fig.~\ref{fig:7}(a)-(b).  However, even in the unstable case they do not ever superimpose to make a sharp peak.  This is due to the conservation of phase quantum number by the NLSE.  For the BEC, this is an important point.  In higher dimensionality solutions to the NLSE with attractive nonlinearity collapse.  This occurs when the density becomes large enough for the nonlinear term to dominate the kinetic energy.  A highly excited, intrinsically complex stationary state, though unstable, can lower the maximum density by a factor of forty or more, thereby preventing collapse.

Finally quintic nonlinearity is added, as in Eq.~(\ref{eqn:nlsequintic}).  For $b=0.1$ the cn-type stationary state in the adjacent regime pulses but is otherwise unaffected, as may be seen in Fig.~\ref{fig:7}(c).  In Fig.~\ref{fig:7}(d) it is seen that for $b=-0.1$ the same stationary state  undergoes large oscillations but does not lose its integrity as a soliton-train.  Although not illustrated here, the ground state at this scale, a single density peak, deforms slightly but is otherwise unaffected for $b=\pm 0.1$.

\section{Harmonic potential}
\label{sec:harmonic}

Since many BEC experiments involve 
magneto-optical 
traps~\cite{ketterle1}, the consideration of the stability properties of the stationary states of such a potential has immediate experimental relevance.
A harmonic confining potential is added to Eq.~(\ref{eqn:nlsetime}) and it is rescaled to the form
\begin{equation}
\left[- \partial_{x}^{2}\:\pm\eta\mid\!f(x,t)\!\mid^{2}
   + ax^2\,\right]\,f(x,t) = \imath\partial_t f(x,t),
\label{eqn:nlseharmonic}
\end{equation}
where $a$ measures the strength of the harmonic trap and $\eta$ the strength of the nonlinearity, which is proportional to the number of condensed atoms.  As before, we consider the class of stationary solutions
to Eq.~(\ref{eqn:nlseharmonic}) by letting 
$f(x,t)=f(x)e^{-i\mu t}$, so that
\begin{equation}
\left[- \partial_{x}^{2}\:\pm\eta\mid\!f(x,t)\!\mid^{2}
   + ax^2\,\right]\,f(x) = \mu f(x)
\label{eqn:harmoniceig}
\end{equation}
is the resulting eigenvalue problem with eigenvalue $\mu$.  

\subsection{Stability of stationary states}
\label{ssec:harm}

\begin{figure}
\begin{center}
\epsfig{figure=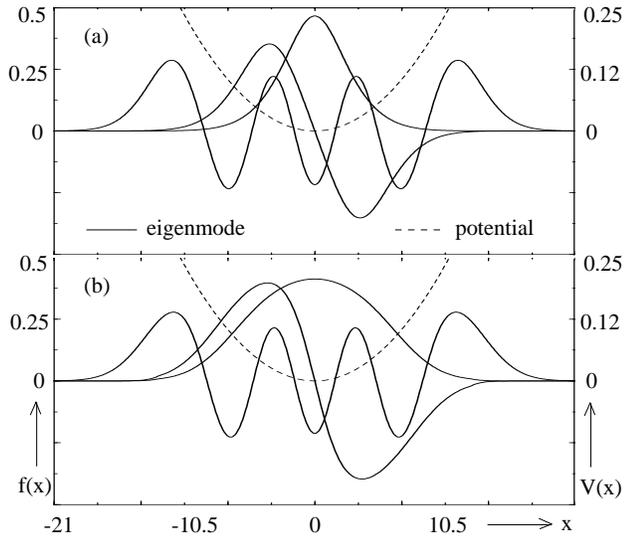,width=8.2cm}
\end{center}
\caption{
Stationary states of the harmonic potential for (a) attractive and (b) repulsive nonlinearity.  Pictured are the ground state, first mode, and sixth mode.  Note that states in the overlapping regime, such as the sixth mode, are predominantly linear, while those in the adjacent regime, such as the first mode, are strongly nonlinear.  There is no well-separated regime in the harmonic potential.
}
\label{fig:8}
\end{figure}

The addition of the harmonic potential no longer allows closed
form solutions to be obtained.  However, the normalized
eigenfunctions of Eq.~(\ref{eqn:harmoniceig}) can be constructed
numerically via standard shooting methods~\cite{kivshar6}.  
Fig.~\ref{fig:8} depicts the stationary solutions
of Eq.~(\ref{eqn:harmoniceig}) normalized to unity with $a=0.02$ and 
$\eta=1$.  
In Fig.~\ref{fig:8}(a)-(b) the zeroth, first, and
sixth modes are depicted for attractive and repulsive nonlinearity, 
respectively.  In the lower modes it is seen that attractive 
nonlinearity sharpens the peaks and troughs 
while repulsive nonlinearity makes them spread out.
Highly excited states, as for example the sixth mode, are further in the linear regime, and therefore resemble the Hermite 
polynomial stationary solutions to the analogous problem in 
linear quantum mechanics.  These results are qualitatively in accord with
the sn-type and cn-type closed form~\cite{carr15,carr16} stationary states for box or periodic boundary conditions.

\begin{figure}
\begin{center}
\epsfig{figure=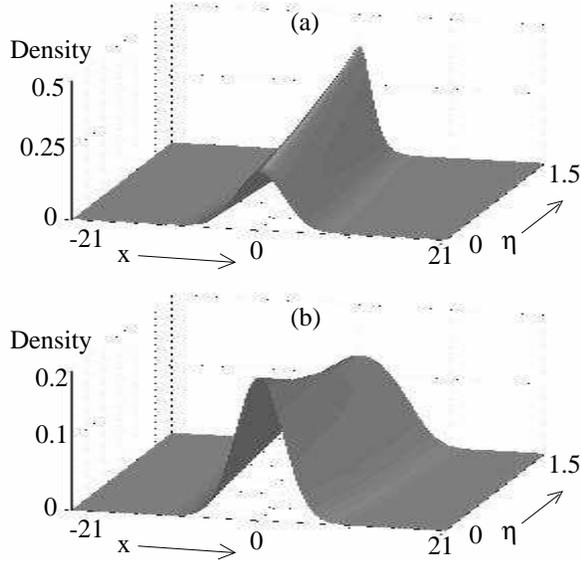,width=8.2cm}
\end{center}
\caption{
Dependence of the ground state on the strength of the nonlinearity for the harmonic potential.  (a) For attractive nonlinearity the ground state deviates rapidly from the linear gaussian solution towards a sharp peak.  (b) For repulsive nonlinearity it broadens to become parabolic with gaussian tails.}
\label{fig:9}
\end{figure}

As the condensate is grown the effective number of atoms is
increased, thereby increasing the strength of the nonlinearity
via the parameter $\eta$.  To illustrate this, Fig.~\ref{fig:9}
depicts the density of the zeroth mode as a function of increasing nonlinearity.  
Here $\eta$ varies from 0 to 3 and the eigenfunctions are clearly seen to 
deform as $\eta$ increases.
For attractive nonlinearity, the
ground state steepens significantly from the 
original Gauss-Hermite polynomial which is the solution for
the linear case of $\eta=0$.  In contrast, repulsive nonlinearity 
causes the Gauss-Hermite polynomial to spread out as the nonlinearity is increased.  It takes on a parabolic shape with Gaussian tails, known as the Thomas-Fermi limit for the BEC.  Thus as atoms are condensed in a harmonic potential trap, the effect of the
mean-field contribution is to significantly reshape the stationary states.

To study the stability of these states a substantial initial white-noise perturbation was added and then Eq.~(\ref{eqn:nlseharmonic}) was solved.  The zeroth through sixth modes were tested, but for the sake of brevity only 
the first and sixth ones are illustrated.  In accord with the results of Sec.~\ref{sec:pbc} it was expected that all solutions would be stable for repulsive nonlinearity, while for attractive nonlinearity they would be stable in the adjacent but not the overlapping regime.  Note that here there is no well-separated regime, due to the effect of the harmonic potential.

\begin{figure}
\begin{center}
\epsfig{figure=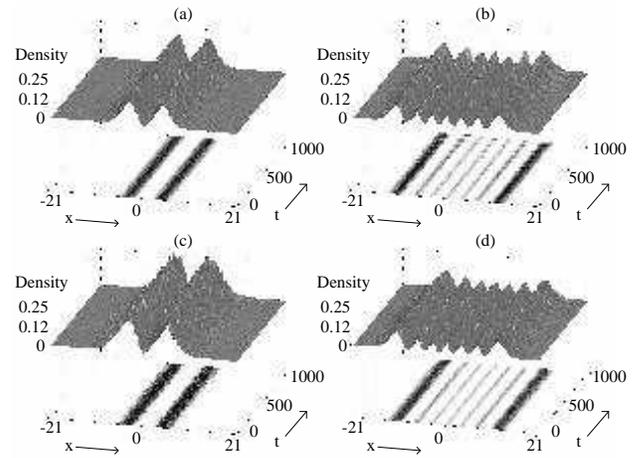,width=8.2cm}
\end{center}
\caption{
Stability of stationary states in the harmonic potential.  The first and sixth modes are propagated with initial white noise over long time scales for (a)-(b) attractive nonlinearity and (c)-(d) repulsive nonlinearity.  The stability properties are identical to those of the analogous solutions for periodic solutions on the ring: only (b) shows instability, as it is for attractive nonlinearity in the overlapping regime.
}
\label{fig:10}
\end{figure}

Fig.~\ref{fig:10}(a)-(b) and (c)-(d)
depicts the results of the simulations for $a=0.02$ and $\eta=1$ for a time of $t=1000$ for attractive and repulsive 
nonlinearity, respectively.  Both the perturbed attractive  
and repulsive
stationary states are observed to be stable over very long times.  
Figure~\ref{fig:10}(b) however, starts to exhibit an oscillatory
instability at long times, much like that observed for the
overlapping regime in Sec.~\ref{sssec:att}.

Although not illustrated here, the effect of either negative or positive quintic nonlinearity of the form and magnitude described in Sec.~\ref{sssec:rep} is to slightly deform the eigenstates but leave them otherwise unaffected.

\subsection{Phase engineering}
\label{ssec:manip}

Phase engineering has been used to successfully create solitons in BECs confined in a harmonic potential~\cite{denschlag1,burger1}.  Specifically, a step function in the phase is used to create a density notch which then propagates across the condensate.  It has been shown elsewhere~\cite{carr22,reinhardt1} that in quasi-one-dimensional confinement both bright and dark solitons may be manipulated, or phase engineered, by various simple phase profiles.  Here it is shown that the same is true of excited states in a harmonic potential.

To induce dynamics in the stationary states, the
eigenfunction solutions depicted in Fig.~\ref{fig:8} were modified by
introducing the following two phase profiles into the initial
conditions:
\begin{eqnarray}
  &&  \mbox{I.} \,\,\,\,\, f(x,0)=f(x)\exp(i\beta x) \\
  &&  \mbox{II.} \,\,\,\,\, f(x,0)=f(x)\exp(i\beta|x|) \, ,
\end{eqnarray}
where $\beta$ determines the phase pertubation strength.  
For profile I, all peaks are ramped in the same
fashion, whereas for profile II, the peaks
are ramped in opposite directions initially depending
upon their location in $x$.  

\begin{figure}
\begin{center}
\epsfig{figure=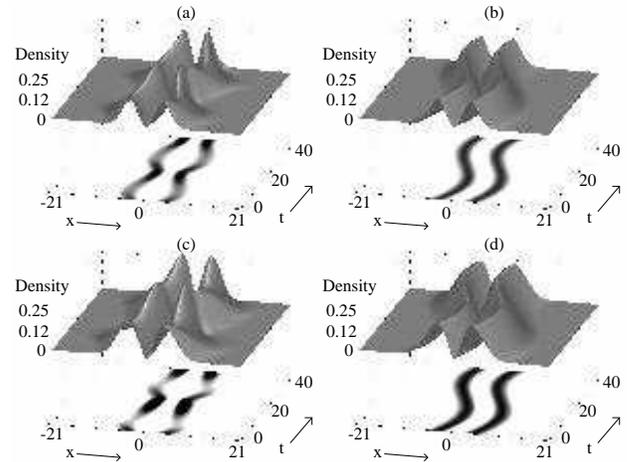,width=8.2cm}
\end{center}
\caption{
Phase engineering of stationary states in the harmonic potential.  The particle-like nature of these solutions is exhibited for both (a)-(b) attractive nonlinearity and (c)-(d) repulsive nonlinearity.  In (a) and (c) an equal and opposite linear phase ramp was used on the two density peaks, while in (b) and (d) an identical phase ramp was used.  This produces two oscillatory modes analogous to those of the coupled pendulum.
}
\label{fig:11}
\end{figure}

Fig.~\ref{fig:11} illustrates the resulting dynamics for both attractive and repulsive nonlinearity.  In (a) and (c), phase profile II is imposed
on the first mode for $a=0.02$, $\eta=1$, and $\beta=0.3$.  
This initially leads to a repulsion
of the peaks.  However, the potential counteracts this effect
and the peaks undergo an oscillatory particle-like motion.  
In contrast, phase profile I keeps the peaks moving in unison and the
potential once again acts to trap the peaks inside the potential
by generating a periodic motion within the well.  These two cases are 
analogous to the oscillatory eigenmodes of the coupled pendulum,
and clearly demonstrates the soliton-like behaviour of the stationary states when phase engineered.  Thus in the mean-field approximation the BEC, itself consisting of particles, has solutions of a particle-like nature.

\section{Conclusion}
\label{sec:conclusion}

The stability of excited stationary states of a trapped gaseous
Bose-Einstein condensate in a quasi-one-dimensional environment has been
studied with respect to addition of stochastic white noise, externally imposed phase profiles (phase engineering), and perturbations of the effective
Hamiltonian by inclusion of interactions proportional to higher powers of
the density.  This study has been carried out within the context of
investigating  perturbations to the simplest mean-field picture: that
which gives rise to the cubic nonlinear Schr\"odinger equation.  Such
excited states are soliton-like, and, as such, have recently
been observed experimentally for the 
BEC~\cite{denschlag1,burger1}.  

Two topologies were considered: periodic solutions on a ring and confinement in a harmonic potential.  It was found that in both physical contexts all of the above types of perturbations basically left the solitonic structures intact, even when the initial stationary states had become unstable.  In quasi-one-dimension, harmonic confinement enhanced stability.  For attractive nonlinearity the perturbed stationary states were either stable or quasi-periodically stable.  The most stable solution-type was one for which alternating peaks were placed adjacent to one another and separated by nodes~\cite{desem1,michinel1}.  For repulsive nonlinearity, excited states drifted but were otherwise stable.


\acknowledgments

We benefited greatly from discussions with David Thouless and Mary Ann Leung.  Sarah McKinney and Joachim Brandt provided additional computational assistance.  This work was supported by the National Science Foundation Grants CHE97-32919 for L. D. Carr and W. P. Reinhardt and DMS-9802920 for J. N. Kutz.



\begin{references}
\bibitem[*]{byline} to whom correspondence should be addressed
\bibitem{whiteNoise}
White noise is included by adding to every Fourier
mode a Gaussian distributed random number with
mean zero and unit variance multiplied by a strength
coefficient.  The strength coefficient utilized here
is typically 0.5 divided by the number of Fourier modes.
\end{references}

%
%

\pagebreak

\end{document}